\documentclass[10pt,prb,superscriptaddress,nofootinbib,showpacs]{revtex4}

\usepackage{graphicx}
\begin{document}

\title{Geometrical and statistical factors in fission of small metal clusters}

\author{O. I. Obolensky} 
\email{oleg@mail.ioffe.ru}
\affiliation{Frankfurt Institute for Advanced Studies, Johann Wolfgang Goethe-University,
Robert-Mayer Strasse 10, D-60054 Frankfurt am Main, Germany}
\affiliation{A.F. Ioffe Institute, Politechnicheskaja str. 26,
St. Petersburg 194021, Russia}

\author{A. G. Lyalin}
\affiliation{Institute of Physics, St. Petersburg State University, Ulianovskaja 1,
Petrodvorez, St. Petersburg 198504, Russia}

\author{A. V. Solov'yov}
\altaffiliation[On leave from: ]{A.F. Ioffe Institute, Politechnicheskaja str. 26,
St. Petersburg 194021, Russia}
\email{solovyov@fias.uni-frankfurt.de}
\affiliation{Frankfurt Institute for Advanced Studies, Johann Wolfgang Goethe-University,
Robert-Mayer Strasse 10, D-60054 Frankfurt am Main, Germany}

\author{W. Greiner}
\affiliation{Frankfurt Institute for Advanced Studies, Johann Wolfgang Goethe-University,
Robert-Mayer Strasse 10, D-60054 Frankfurt am Main, Germany}

\begin{abstract}
Fission of metastable charged univalent metal clusters has been
studied 
on example of $Na_{10}^{2+}$ and $Na_{18}^{2+}$ clusters
by means of density functional theory methods.
Energetics of the process, i.e. dissociation energies
and fission barriers, 
as well as its dynamics, i.e. fission pathways, 
have been analyzed.
The dissociation energies and fission barriers
have been calculated
for the full range of fission channels for the $Na_{10}^{2+}$  cluster.
Our data confirms the viewpoint
that there is some correlation between these two quantities,
which can often be explained by electronic shell effects.
However, there is no quantitative correspondence between the
dissociation energies and heights of fission barriers
and the former can not serve as a quick estimate
for the latter.
The impact of cluster structure on the fission process
has been elucidated.
The calculations show that the geometry of the smaller fragment
and geometry of its immediate neighborhood in the larger fragment
play a leading role in defining the fission barrier height:
energy barriers for 
removing different combinations of atoms
from the same parent cluster can vary greatly, 
while energy barriers for
removing similar groups of atoms
from similar places in different isomers of the parent cluster are 
usually similar. 
The present study demonstrates importance of
rearrangement of the cluster structure
during fission.
The rearrangment can lower the fission barriers significantly.
It may include forming a neck between the two fragments 
or fissioning via another isomer state
of the parent cluster;
examples of such processes are given.
For several low-lying isomers of $Na_{10}^{2+}$ cluster
the potential barriers for transitions between these isomer states
are calculated and compared with the corresponding fission barriers.
These data suggest that 
there is a competition between
``direct" fission 
and fission going via intermediate isomer states
of the parent cluster.
An impact of the cluster geometry on the
change of the system's entropy due to fission is 
also discussed.
\end{abstract}


\maketitle

\section{Introduction}

Studying properties of charged clusters is of great interest
both from theoretical and experimental viewpoints.\cite{ISACC03}
In experiments, widely used clusters detection techniques
proceed via ionization and subsequent mass spectrometry of clusters
(see Refs.~\onlinecite{LesHouches,deHeer_RMP93} 
for a review and references).
On the other hand, charged clusters behaviour 
presents
one of the instances of a long standing fundamental
theoretical problem of stability of complex systems
since charged clusters can be found in stable,
metastable or unstable states, depending on cluster size
and excessive charge.
Metal clusters are especially attractive
subject of study because fission of charged metal clusters 
provides close analogies to
the corresponding processes in nuclear systems.

For semi-quantitative classification of stability of charged
metal clusters it is convenient to use a fissility
parameter introduced by Lord Rayleigh
more than a century ago,\cite{Rayleigh_PM1882}
while he was investigating
stability of small charged liquid droplets.
The fissility parameter is defined as ratio
of the Coulomb to surface energies of the drop,
$X = E_{\rm Coul}/2E_{\rm Surf}$.
For spherical sodium clusters\cite{Guet_PRL01}
$X \approx 2.5 \, Q^2/N$
($Q$ and $N$ are the cluster charge and size, respectively).
For $X \ll 1$ clusters are stable, for
$X \gg 1$ clusters are unstable and fragment
by the Coulomb explosion.
For $X \sim 1$ clusters are found in metastable states and
must overcome a potential barrier in order to fission.
As in many other recent 
studies,\cite{Guet_PRL01,Brechignac_PRL04,LandmanBrechignac_PRL02,%
Wang_PRL02,BlaistenBarojas_PRB98,Guet_PRL95,Brechignac_JCP94} 
of both experimental and theoretical character,
we focus on the latter situation and present results
of our 
all-electrons density functional calculations
which help
to achieve better understanding of various aspects of fission 
of metastable charged metal clusters.
We use sodium clusters, well studied experimentally, as
a sample revealing general properties of univalent metal clusters.

Within the Born-Oppenheimer approximation,
we employ density functional theory methods for calculating
the electronic 
structure
for every given configuration of the nuclei. Our approach
differs from the standard 
technique of molecular dynamics simulations.
In our calculations we explore the multi-dimensional potential 
energy surface of a cluster system in order 
to find the local minima on this surface, corresponding to
different cluster isomers. The potential energy surface also
allows to determine the optimal fission pathways which minimize
the fission barriers. 

We begin with calculating 
the energetics of the fission process. 
We present the dissociation energies and fission barriers
for all possible fission channels for the $Na_{10}^{2+}$ cluster
for which the excessive charge is distributed between
the daughter fragments,
$Na_{10}^{2+} \to Na_{P}^{+} + Na_{N-P}^{+}$, $P=1..5$.
Such channels are energetically more favourable compared
to the evaporation channels in which a neutral and double
charged fragments appear in the final state.

The dissociation energies is
an important characteristic of the fission process,
useful in assessing stability/instability
of a particular cluster. 
Comparison of the dissociation energies and fission barriers
shows
that there is 
some correlation between the two quantities.
This correlation can often be
explained by electronic shell effects which favour forming
fragments with filled and half-filled shells.
However, 
as one may expect,
the correlation is only qualitative,
and therefore dissociation energies can not serve
as an easy-to-calculate, convenient tool for making 
reliable predictions concerning the fission process 
(e.g. predictions of the preferred fission channel 
or branching ratios between different channels) and
one needs to find fission barrier heights.

On the other hand, even knowledge of the fission barrier heights
is sometimes not sufficient for predicting
the preferred fission channel.
This is the case for the dominant asymmetric,
$Na_{10}^{2+} \to Na_{7}^+ + Na_3^{+}$,
and symmetric,
$Na_{10}^{2+} \to 2 Na_{5}^+$, 
fission channels for the $Na_{10}^{2+}$ cluster.
Considerations based on electronic shell effects
suggest that these two channels ought to be
the favoured ones. The accurate {\it ab initio}
calculations confirm this conclusion, but it turns
out that the barrier heights for these channels 
are so close that one needs to take into
account geometrical and statistical factors,
as discussed below (section \ref{subsection-geometry}).

Our approach permits also
studying the dynamics of the fission process. 
Having calculated the multi-dimensional energy surface
we were able to determine the optimal dynamic
pathways of the fission for all the channels considered.
Analysis of the data suggests that
significant rearangement of the cluster structure often
accompanies fission.

One can distinguish between two main types of
such rearrangement.\cite{Rearrangement_JPB04} 
The first one is a rearrangement
of the cluster structure without significant change
in distance between the centers of mass of the prospective
fragments. This type of rearrangement takes place
before the actual separation of the fragments begins
and it is, in fact, a transition to another isomer
state of the parent cluster.\cite{Rearrangement_JPB04}
The second type of rearrangement is characterized
by existence of a super-molecule-like intermediate state
in which the fragments are sufficiently separated
from each other, but connected 
by a "neck".\cite{Rearrangement_JPB04,Montag_PRB95,LandmanBrechignac_PRL94,Landman_PRL91}
A similar necking phenomenon 
is known for nuclear fission.\cite{EisembergGreiner_85}
Necking allows for great
reduction in the overall height of the fission barrier
and is responsible for a double humped form of
the barrier.
In dynamical simulations 
necking can be observed
as an elongation of the cluster shape
during fission.\cite{Guet_PRL01,BlaistenBarojas_PRB98}

The first type of rearrangement implies that
there is a competition between one-step
(``direct" fission) and two-step (fission 
via an intermediate
isomer state) processes. Therefore,
in order to make predictions concerning
the fission activation energy and fission pathway
for a given cluster isomer
one has first to find the energy barrier for the direct fission
of the given isomer and then to compare the height
of this barrier with the heights of the fission barriers
for other low lying isomers. The differences in the
isomer energies and potential barriers between the
isomers have also to be taken into account.
We present such analysis for the $Na_{10}^{2+}$ cluster.

Accounting for geometry of a fissioning
cluster\footnote{This is, of course, sensible for cluster
temperatures below the melting point only. 
For temperatures above the melting point
the atoms in clusters do not have stationary positions
and one can not speak of a certain geometry of a cluster.}
necessitates an extra care in calculating the fission barriers,
since removing different combinations of atoms
requires overcoming different 
barriers.
In other words, different combinations of atoms
are not necessarily equivalent in respect to fission.
Depending on the symmetry group of a cluster
there may exist
the groups of atoms with the same
potential energy barriers.
When searching for the fission barrier of a cluster one needs
to identify all such groups of atoms and choose the group
with the lowest barrier.

It is sometimes also possible to identify
geometrically similar groups of atoms, belonging to
different isomers, for which 
the energy barriers
are very close, even though the overall geometries of
the isomers can be quite unalike.
This suggests that the geometry of the smaller fragment
and geometry of its immediate neighborhood in the parent cluster
play a leading role in defining the energy barrier height.

Accounting for cluster geometry,
which leads to non-equivalence of different
combinations of atoms
may have a further important impact on the way
how the change of the system entropy due to fission should
be calculated.
It has been argued\cite{Brechignac_PRL96,Brechignac_PRL98} that
accounting for an entropy change contribution to the free energy
of the system is necessary for correct
description of the temperature and size dependences
of the branching ratios between different fission channels,
while purely energetic considerations based on fission barrier
heights fail. 
The entropy change is, in fact, a change in the statistical
weight of the initial and final states of the system.
In order to calculate the change in the statistical weight 
one has to count the change in number of
combinations out of which the initial and final states
of the system are composed. The non-equivalence of
different combinations of atoms leads to replacement
of the binominal law, used when all the atoms in a cluster are
equivalent,\cite{Brechignac_PRL98} by a more accurate treatment,
which has to take into account particular geometry of a given isomer.
We base such a treatment on counting the number of 
equivalent combinations of atoms with
the minimum potential energy barrier.
The type of statistics to be used when calculating 
the change of the system's entropy is determined by the
cluster temperature. At high temperatures
the cluster is melted and does not
posess a certain structure. In this situation the purely
combinatorial approach is fully justified.
Conversely, at low temperatures a more detailed
account of the cluster structure is needed.

We note, that the kinds of studies mentioned above
are beyond the scope of simpler approaches
which do not take into account ionic structure of clusters.
In such approaches, 
e.g. in the jellium model, one may speak
of, say, prolate and oblate  jellium shapes
as of different isomers, but this does not reflect
all the variety of energies and geometries of
stable cluster isomers.

\section{Simulation details}

\subsection{General formalism}
We utilize the methods
of density functional theory (DFT) within the Born-Oppenheimer scheme.
In accord with the DFT prescriptions
we iteratively solve the Kohn-Sham equations\cite{KohnSham_PRA65}
\begin{equation}
\left(
\frac{p^2}{2}+U_{\rm i}+V_{\rm H}+V_{\rm xc}
\right)
\psi_i = \varepsilon_i \psi_i,
\label{KSeq}
\end{equation}
\noindent where the first term corresponds to the kinetic
energy of an electron,
$U_{\rm i}$ describes the attraction of the $i^{\rm th}$ electron to the nuclei
in the cluster, $\psi_{\rm i}$ is the electronic orbital,
$V_{\rm H}$ is the Hartree part of the
inter-electronic interaction,
\begin{equation}
V_{\rm H}({\bf r})=\int
\frac{\rho({\bf r}^\prime)}{|{\bf r}-  {\bf r}^\prime |}
{\rm d} {\bf r}^\prime ,
\end{equation}
\noindent $\rho({\bf r})$ is the electron density,
$V_{\rm xc}$ is the local exchange-correlation potential
defined as the functional derivative
of the exchange-correlation energy functional
\begin{equation}
V_{\rm xc} = \frac{\delta E_{\rm xc} [\rho]}{\delta \rho({\bf r})},
\end{equation}
\noindent where the exchange-correlation energy is partitioned
into two parts, referred to as exchange and correlation
parts:
\begin{equation}
E_{\rm xc} [\rho]= E_{\rm x}(\rho) + E_{\rm c} (\rho).
\end{equation}

There is a variety of exchange-correlation functionals
in the literature.
We have used the three-parameter Becke-type gradient-corrected
exchange functional with the gradient-corrected correlation
functional of Lee, Yang, and Parr (B3LYP).
For the explicit form of this functional we refer to the original papers
\cite{Becke_PRA88,Becke_JCP93,LYP_PRB88}.
The B3LYP functional has proved to be a reliable tool for
studying the structure and properties of small sodium clusters.
It provides high accuracy at comparatively low computational
costs. For a discussion and a comparison with other approaches,
see Refs.~\onlinecite{Na_structures_PRA02,Mg_structures_PRA03}.

\subsection{Simulation procedure}

We employ a procedure somewhat different
from the usual 
molecular dynamics simulations
techniques. 
In our calculation we explore the 
multi-dimensional potential energy surface of the cluster.
For each point on this surface
we solve Eq.~\ref{KSeq} for the corresponding
geometry of the atomic nuclei
by expanding the cluster orbitals into the basis sets of
primitive Gaussian functions\cite{Chemistry}
with the use of the GAUSSIAN 03 software package.\cite{Gaussian03}

The 6-31G(d) and LANL2DZ basis sets of the Gaussian functions
have been used.
The 6-31G(d) basis set has been used for simulations involving
the $Na_{10}^{2+}$ cluster. This basis set
expands
all the atomic orbitals, so that the dynamics of all particles in the system
is accounted for.
For the $Na_{18}^{2+}$ cluster we have used the more numerically efficient
LANL2DZ basis, for which valent atomic electrons move in
an effective core potential (see details in Ref.~\onlinecite{Chemistry}).
The accuracy and consistency of the calculations is proved by
the correct asymptotics of the total energy of the system at large 
separation distances,
i.e. the total energy of the system at large separation distances 
equals to the sum of the total energies of the charged isolated fragments 
(calculated separately) and the Coulomb repulsion energy.
For certain fission channels we have also compared the fission
barriers obtained with the use of different basis sets (6-311G(d), 6-31G(d),
LANL2DZ) and confirmed that our results do not depend on the choice
of the basis set.

The global minimum on thus found multi-dimensional potential energy surface 
corresponds to the
energetically preferred state of the system.
In the case of metastable doubly charged clusters the global minimum 
corresponds to the system, fragmented into two charged parts.
The fission channel corresponding to the global minimum can be
determined from the dissociation energies which can be generally defined as
\begin{equation}
D_{N,P}^{Q,Q^\prime} = E_{P}^{Q^\prime} + E_{N-P}^{Q-Q^\prime} - E_{N}^{Q},
\label{DNP}
\end{equation}
\noindent where $N$ and $Q$ are the size and the charge
of the parent cluster, $P$ and $Q^\prime$ are the size and the charge
of one of the daughter fragments,
$E_{M}^{W}$ is the total energy of a cluster
of size $M$ and charge $W$.
The global minimum is located 
in the domain of
the potential energy surface
where the distance between the two fragments is infinitely large.
There are other local minima 
at infinitely large distances between the fragments
(at the "edges" of the potential energy surface)
corresponding to other possible fission channels.
The deepest local minimum on the potential energy surface 
which is
located in the "center" part of the surface,
where the two parts of the system are close,
corresponds to the ground state of the cluster, 
while other local minima in the center part of the surface
represent other (meta)stable isomer states.

In this potential energy surface approach the simulation 
of the fission process comes to finding
a pathway on the system's multi-dimensional potential energy surface 
from a minimum in the center part of the surface to a minimum
at its edge.
The found pathway must minimize the energy barrier
for the transition.

In simulation of the fission process we start from 
the optimized geometry of a cluster 
(for details of the geometry optimization procedure 
see Ref.~\onlinecite{Na_structures_PRA02})
and choose the atoms the 
resulting fragments would consist of. 
The atoms chosen for a smaller fragment are shifted 
from their optimized locations in the parent 
cluster to a certain distance. Then, the multi-dimensional potential 
energy surface, its gradient and forces with respect to the 
molecular coordinates are calculated. These quantities specify 
the direction along the surface in which the energy decreases 
most rapidly and provide information for the determination 
of the next place for placing the atoms.
If the fragments are 
removed not far enough from each other then the cohesive 
forces prevail over the repulsive ones and the fragments 
stick together forming the unified cluster again. 
Forming the unified cluster
does not necessarily mean returning to the same point
on the potential energy surface; 
it may happen that the system gets
into another local minimum. This would correspond
to changing the isomer state of the cluster.
Correspondingly, one can find a potential barrier for such transition.
If the fragments are far enough from each other,
the repulsive forces dominate and the 
fragments drift away from each other. The dependence of 
the total energy of the system on the fragments separation 
distance forms the potential energy barrier for a given
pathway. 
Hence, finding the fission barrier is equivalent
to finding the pathway with the lowest
potential energy barrier.

Determining fission barriers is a computationally demanding task
which requires a lot of computer resources.
Currently, it is not feasible to study the multi-dimensional potential energy
surface in fine details even for relatively small clusters.\footnote{For example, 
a calculation of one point of the system's potential energy surface
for the $Na_{10}^{2+}$ cluster takes about a minute on a Pentium Xeon processor
if only valent electrons are considered and about twenty minutes
if all 108 electrons are accounted for.}
Therefore, there is no guarantee that the found fission pathways
provide the lowest fission barriers. Even small deviation
from the found pathway may result in a slight decrease in the fission
barrier height.
We estimate accuracy of the fission barrier heights presented
in this paper to be 0.02-0.04~eV. Further calculations could
allow to obtain a more accurate profile of the fission barrier
and to determine the fission barrier height more precisely. 
However, the accuracy of theoretical models used nowadays 
to describe the experimental results, as well as the accuracy
of the experimental data itself, make such refinements
unneeded.

There is also a possibility that we have overlooked a completely
different fission pathway which would make the fission barrier
even lower. In fact, this was the case with the fission barrier
for the dominant asymmetric fission channel for the
$Na_{10}^{2+}$ cluster, $Na_{10}^{2+} \to Na_{3}^+ + Na_7^+$.
Our previous calculations\cite{Rearrangement_JPB04} had resulted in the fission barrier
of about 0.5 eV for this channel. This was in a good agreement
with the results of other molecular dynamics simulations.\cite{Guet_PRL01,Landman_PRL91}
However, a more detailed study of the multi-dimensional potential energy
surface performed in this work
which included a more complete analysis
of various possible rearrangements of the cluster structure in the course
of fission allowed us to find a lower fission barrier for this
channel, equal to 0.34~eV.

\section{Results and discussion}

\subsection{Fission energetics: dissociation energies and fission barriers}
\label{energetics}

The energetics of the fission process
is characterized by the dissociation energies $D_{N,P}^{Q,Q^\prime}$
and the fission barriers $B_{N,P}^{Q,Q^\prime}$
(here $N$ and $Q$ are the size and the charge of the fissioning cluster
and $P$ and $Q^\prime$ denote the size and the charge of one of
the fragments). 
For metastable clusters, fission barrier  
is an important characteristic
of the fission process, while dissociation energy plays a smaller role.
The parameters of the fission barriers 
largely define many experimentally observable characteristics of the process,
including branching ratios between different fission channels,
fission time, etc.
The dissociation energies are of limited use for metastable clusters.
Their negative signs show that the clusters
are unstable in respect to the corresponding fission channels.
One could hope that easy-to-calculate dissociation energies 
could serve as quick qualitative estimates of fission barriers,
since calculations of fission barriers are much more laborious
and require more computer resources.

\begin{figure}[ht]
\begin{center}
\includegraphics{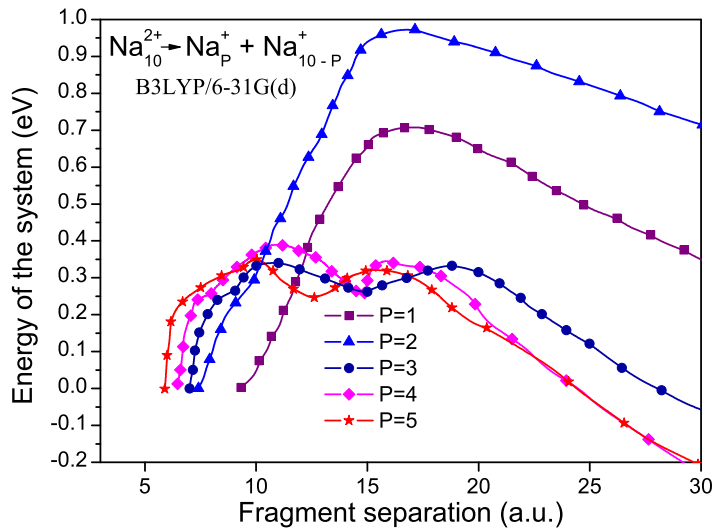}
\end{center}
\caption{(Color online) The fission barriers  
for the different channels of
fission of the $Na_{10}^{2+}$ cluster,
$Na_{10}^{2+} \to Na_{P}^+ + Na_{10-P}^+$.
The zero level of energy is chosen equal to
the energy of the ground isomer state (with distorted $T_d$ symmetry,
denoted in the paper $T_d^{(1)}$) 
of the cluster. The initial
separation distances correspond to the distances between the
centers of masses of the prospective fragments and, consequently,
are finite, so that the barriers do not start at the origin.
}
\label{Na10barriers}
\end{figure}

To answer this question, we have determined the
dissociation energies and fission barriers
for the full range of fission channels 
for the $Na_{10}^{2+}$ cluster,
\begin{equation}
Na_{10}^{2+} \to Na_{P}^{+} + Na_{N-P}^{+}, \qquad P=1..5.
\end{equation} 

\begin{figure}[ht]
\begin{center}
\includegraphics{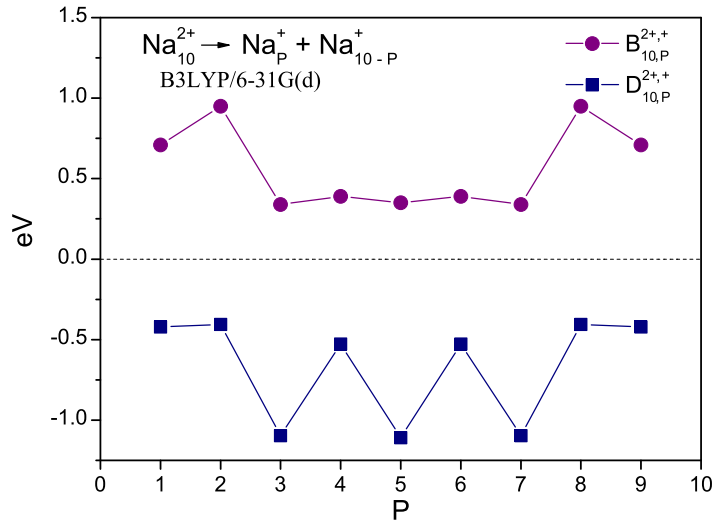}
\end{center}
\caption{(Color online) The fission barrier heights $B_{10,P}^{2+,+}$ (circles) and dissociation
energies $D_{10,P}^{2+,+}$ (squares) 
as functions of the fragment size $P$
for the fission of the ground state
of the $Na_{10}^{2+}$ cluster,
$Na_{10}^{2+} \to Na_{P}^+ + Na_{10-P}^+$.
}
\label{Na10bd}
\end{figure}

The obtained fission barriers are plotted in Fig.~\ref{Na10barriers}.
The fission barrier heights and the dissociation energies
as functions of fission channel are presented in 
Fig.~\ref{Na10bd} and also summarized in Table~\ref{Na10bdTab}.

\begin{table}
\caption{\label{Na10bdTab} 
The dissociation
energies $D_{10,P}^{2+,+}$
and fission barrier heights $B_{10,P}^{2+,+}$ (in eV)
for various fragment sizes $P$
for the fission 
of the $Na_{10}^{2+}$ cluster,
$Na_{10}^{2+} \to Na_{P}^+ + Na_{10-P}^+$.
}
\begin{ruledtabular}
\begin{tabular}{@{}lrrrrr}
                              & $\quad P=1$ & $\quad P=2$& $\quad P=3$ & $\quad P=4$ & $\quad P=5$  \\
\hline 
$D_{10,P}^{2+,+}$ (this work) & -0.38 & -0.37 & -1.07 & -0.52 & -1.11 \\
$B_{10,P}^{2+,+}$ (this work) &  0.71 &  0.97 &  0.34 &  0.39 &  0.35 \\
$B_{10,P}^{2+,+}$, (Ref.~\onlinecite{Rearrangement_JPB04})%
                              &       &       &  0.49 &       &        \\
$B_{10,P}^{2+,+}$, (Ref.~\onlinecite{Guet_PRL01})%
                              &       &       &  0.52 &       &  0.48  \\
$B_{10,P}^{2+,+}$, (Ref.~\onlinecite{Montag_PRB95})%
                              &  0.69 &       &  0.67 &       &       \\
$B_{10,P}^{2+,+}$, (Ref.~\onlinecite{Landman_PRL91})%
                              &  1.03 &       &  0.71 &       &       \\
$B_{10,P}^{2+,+}$, (Jellium model\cite{HFLDA2_PRA02})%
                              &       &       &  0.16  &      &        \\
\end{tabular} 
\end{ruledtabular}
\end{table}

Unfortunately, although there is some correlation in behaviour of
the two functions,
there is no direct correspondence between them.
Therefore, the dissociation energy
can not serve as a quick estimate for the fission barrier height
and one needs to carry out the full calculation
in order to find the fission barrier. Only after that
one can make reliable predictions concerning the fission process.

The correlation in dependence of the dissociation energy and fission
barrier height on the size of the daughter fragment can be
explained by electronic shell effects which favour forming
fragments with filled and half-filled shells.
Indeed, 
the barrier maxima 
are located at distances comparable to or exceeding
the sum of the resulting fragments radii, that is
not far from the scission point.
At such distances the interaction between the fragments,
apart from Coulombic repulsion,
is mainly determined by the electronic properties 
of the system (and also by the cluster geometry in the immediate
vicinity of the scission point),
rather
than by the details of the ionic structure of the bulk
of the fragments.

Indeed, the {\it a priori} electronic shell 
considerations suggest that,
for example, for the $Na_{10}^{2+}$ cluster two
fission channels, namely, 
$Na_{10}^{2+} \to Na_{7}^+ + Na_3^{+}$ and
$Na_{10}^{2+} \to 2 Na_{5}^+$,
dominate.
Our calculations confirm this conclusion, but it turns
out that the barrier heights for these channels 
are so close that the geometrical and statistical factors
become of primary importance.
We discuss the impact of geometry of the cluster
on the fission process 
in the next subsection.

\subsection{Impact of the cluster geometry on fission process}
\label{subsection-geometry}

Detailed analysis of cluster geometry in the course of fission allows for
deeper understanding of the process. It, however, brings in
two issues which must be properly accounted for.
The first one is the fact that the different combinations of 
equal number of atoms
are not necessarily equivalent for a given cluster geometry,
and the potential barriers to be overcome in order to remove
these combinations of atoms from the parent cluster
are quite likely to differ. The second issue is that not
always ``direct'' fission (i.e. straightforward removing some
combination of atoms from the parent cluster) has lower
potential barrier than fission going via an intermediate isomer state
of the parent cluster or via formation of a super-molecule-like
necked structure. This latter issue will be discussed in detail
in the next subsection.

The problem of finding the lowest fission barrier 
is simplified in many cases
by the presence of a symmetry in a cluster geometry which reduces the number
of non-equivalent combinations of atoms participating in
the fission process.

\begin{figure}[ht]
\begin{center}
\includegraphics[scale=0.4]{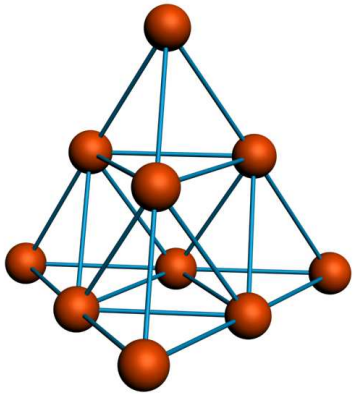}
\end{center}
\caption{(Color online) The geometry of the ground state isomer of
the $Na_{10}^{2+}$ cluster. It has no point symmetries,
but its structure is, in fact, a pyramid 
($T_d$ symmetry group),
slightly distorted due to the
Jahn-Teller effect.
}
\label{Td}
\end{figure}

Let us consider fission of the $Na_{10}^{2+}$ cluster.
Its ground state isomer 
has a slightly distorted symmetry of the $T_d$ point symmetry group\cite{Na_structures_PRA02}
and can be described as a pyramid, see Figure \ref{Td}. 
We will denote this isomer
as $T_d^{(1)}$. Four atoms in the isomer are located in the
vertices of the pyramid and six are in the centers of the edges.
All the edge and all the vertex atoms can be considered equivalent.
Therefore, when considering the fission channel
$Na_{10}^{2+} \to Na_{9}^{+} + Na^{+}$ it suffices
to find two energy barriers only
rather than calculating a barrier for each atom. 
The potential energy barrier
height for removing an atom from an edge of the cluster
is 0.85~eV, while it is 0.71~eV for removing an atom
from a vertex. Hence, the fission barrier height for
the fission channel is 0.71~eV.

There are more energy barriers which need to be calculated
for the $Na_{10}^{2+} \to Na_{8}^{+} + Na_2^{+}$ channel.
Totally, there are 45 combinations of two atoms out of ten.
Fortunately, there are only five energy barriers 
which are to be calculated.
Indeed, two atoms which are to be removed from the parent
cluster can be taken either both from the vertices, or
both from the edges, or one atom from a vertex and one
from an edge. All combinations of two atoms taken from
the vertices are equivalent. There are six such
combinations which we will denote as VV. 
Twenty four combinations of two atoms in which one atom comes
from a vertex and one from an edge can be further divided
into two equal subgroups. In the first subgroup (VE1) 
atoms belong to the same edge while in the second subgroup
(VE2) they do not. Fifteen combinations of atoms from the edges
are also divided into two subgroups. The first one
contains atoms belonging to the same face of the pyramid
(twelve combinations, EE1) and the second group (EE2)
contains the combinations of atoms from the different pyramid
faces.  
The heights of the energy barriers for these five groups
are summarized in Table \ref{groups}.

\begin{table}
\caption{\label{groups} 
The potential energy barrier heights
for removing various combinations of two atoms
from the parent $Na_{10}^{2+}$ cluster, see the text
for explanations.
}
\begin{ruledtabular}
\begin{tabular}{@{}rrrrrl}
$\quad$ VV& $\quad$ VE1& $\quad$ VE2& $\quad$ EE1& $\quad$ EE2& \\
\hline
1.63 & 0.97 & 1.11 & 0.97 & 1.18 & eV
\end{tabular} 
\end{ruledtabular}
\end{table}

One can see from Table \ref{groups} that
in the case of the $Na_{10}^{2+} \to Na_8^+ + Na_2^+$
channel, the two atoms which are to be removed from
the parent cluster in the fission process must
be taken either from a vertex and the middle of an edge
originating from this vertex
or from
the middles of two edges belonging to the the face
of the pyramid.
Other combinations of atoms have higher potential
barriers and do not participate, as a good approximation,
in the process.
In other words, the daughter fragment 
should contain
certain classes of combinations of atoms only,
in order to obtain a lower potential energy barrier.
Our analysis shows that this conclusion is also valid
for other fission channels and parent cluster isomers.
Typically, the combinations which provide the minimum energy barrier
include neighbouring atoms located in a part
of the parent cluster which can be the most easily removed.

\begin{figure}[ht]
\begin{center}
\includegraphics[scale=1]{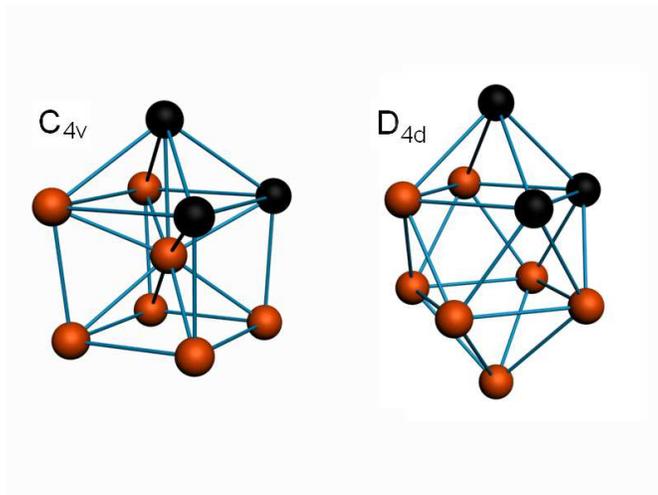}
\end{center}
\caption{(Color online) Two isomers of the $Na_{10}^{2+}$ cluster. From left to right:
an isomer of $C_{4v}$ point symmetry group (nicknamed iso1414);
an isomer of $D_{4d}$ point symmetry group (nicknamed iso1441).
The similar combinations of three atoms which provide very close 
fission barrires are marked with black color.}
\label{isostructures}
\end{figure}

\begin{figure}[ht]
\begin{center}
\includegraphics{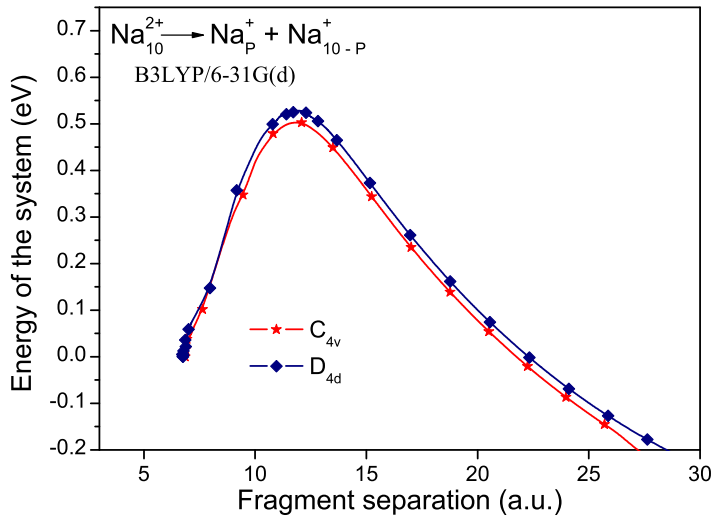}
\end{center}
\caption{(Color online) Energy barriers for  removing 
similar combinations of three atoms 
(marked by black color in Figure~\protect{\ref{isostructures}}) 
from the two different isomers of the $Na_{10}^{2+}$ cluster.
The barriers plotted versus
distance between the centers of mass of the fragments.
The curve with the stars corresponds to
the isomer with the symmetry of $C_{4v}$ point group (iso1414);
the curve with the diamonds corresponds to
the isomer posessing the symmetry of $D_{4d}$ point group (iso1441).
Note, that energies are measured from the energy of the 
initial state of the corresponding isomer, 
i.e. we plot 
$E - E_{C_{4v} (D_{4d})}$, where $E$ is the total energy of the system and 
$E_{C_{4v} (D_{4d})}$ are the energies of the 
iso1414 and iso1441
isomers, respectively. 
}
\label{isobarriers}
\end{figure}

On the other hand, it is sometimes possible to identify
geometrically similar groups of atoms belonging to
different isomers for which 
the energy barriers are very close, 
even though the overall geometries of the isomers
are not similar and can even be quite unalike.
This fact taken in conjunction with the non-equivalence
of different combinations of atoms wihin one isomer state,
suggests that the geometry of the smaller fragment
and geometry of its immediate neighborhood in the larger fragment
play a leading role in defining the energy barrier height.
We can demonstrate this
by comparing energy barriers
for removing similar groups of atoms from similar places
in the parent cluster for two different isomer states
of the $Na_{10}^{2+}$ cluster. 
One of these isomers 
possesses the symmetry of
$C_{4v}$ point symmetry group and
another one is of $D_{4d}$ symmetry.
The isomers geometries are depicted in 
Figure \ref{isostructures},
the groups of atoms to be removed are shown in black color.
The resulting barriers are presented in Figure \ref{isobarriers}.
It is seen from the figure that the barriers are similar not
only in height but also in shape. This can be explained by
similarities of the geometries of the prospective smaller
fragments, similarities in the immediate neighbourhoods 
of the smaller fragments, and also
by similarities in the charge distribution in the isomers, and
by absence of additional structure rearrangements in the course
of fission.

The fact of inequivalence of different combinations
of atoms in a cluster isomer 
may have an important impact on 
various characteristics of the fission process such as
branching ratios between different fission channels
and their temperature dependence.

It has been argued theoretically and shown experimentally that 
dominant fission channels and the branching ratios
between different channels are not governed
by purely energetic considerations but also
by the free energy change which takes into account
the different combinations of atoms which constitute
the fragments.\cite{Brechignac_PRL96,Brechignac_PRL98}
Accounting for an entropy change contribution 
to the free energy of the system was necessary for correct
description of the experimental results.
The branching ratio $I_2/I_1$ between two channels is then defined
by the difference in the fission barrier heights 
$\Delta B \equiv B_{N,P_2}^{Q,Q^\prime} - B_{N,P_1}^{Q,Q^\prime}$
and by the entropy change $\Delta S$ of the system:

\begin{equation}
\frac{I_2}{I_1} = f \exp(\frac{\Delta S}{k} - \frac{\Delta B}{k T}).
\label{branching}
\end{equation}
\noindent Here $k$ is the Boltzmann constant, $T$ is the cluster
temperature, $f$ is a frequency factor.
This factor arises due to different frequencies $\omega$
of oscillations in the modes leading to
fragmentation of the cluster in different fission channels.
Roughly speaking,
the fragmentation occurs when due to a fluctuation 
energy concentrates in a single oscillation mode 
in the amount sufficient for
overcoming the fission barrier.
Fission would be more probable if the frequency
of oscillations in this mode is higher, since then
the system would "attempt to fragment" more frequently.
Hence, in a crude approximation $f \sim \omega_2/\omega_1$.
If one assumes that the potential surfaces for both modes
can be approximated by parabolas with 
same curvatures,
then the ratio of the frequencies of oscillations will depend only
on the reduced masses $\mu$ of the pairs of the prospective fragments,
$f \sim \sqrt{\mu_1/\mu_2}$. For the dominant asymmetric and
symmetric fission channels for the $Na_{10}^{2+}$ cluster
the frequency factor $f$ is about 1.1, so we will neglect
its influence and put $f=1$.
Of course, this factor can be estimated more accurately.
In order to do that one has to determine all the normal
modes of oscillations of the parent cluster, to represent
the oscillations leading to the given fission channel
via a combination of normal mode oscillations and calculate
the probability that enough energy would fluctuate
into these normal modes. Such an analysis 
would go beyond the aims of the current
work and can be a subject of a separate study. 
We now focus on the influence of the cluster
structure on the way how the system's entropy should be
calculated.

The entropy change entering the equation (\ref{branching})
is, in fact, a change in the statistical
weight of the initial and final states of the system.
In order to calculate the change in the statistical weight $\Gamma$
one has to count the change in number of
combinations out of which the initial and final states
of the system are composed,
\begin{equation}
\Delta S = k \ln \Gamma_2 - k \ln \Gamma_1 =
k \ln \frac{\Gamma_2}{\Gamma_1}.
\end{equation}
\noindent For high temperatures 
all atoms can be considered equivalent and the number
of combinations of $P$ atoms out of $N$ is given by
the binominal coefficient, $\Gamma = C_N^{P}$, so that
\begin{equation}
\Delta S = k \ln \frac{N!}{(N-P_2)!P_2!} - k \ln \frac{N!}{(N-P_1)!P_1!}=
k \ln \frac{(N-P_1)!P_1!}{(N-P_2)!P_2!}
\end{equation}

The non-equivalence of
different combinations of atoms requires replacement
of the binominal law, used in Ref.~\onlinecite{Brechignac_PRL98} for high temperature
regime when all the atoms in cluster are
equivalent, by a more detailed treatment,
which has to take into account particular geometry of a given isomer.
Such a treatment should be based on counting the number of 
equivalent combinations of atoms which provide
the minimum potential energy barrier (for low temperatures it is
justified to assume that fission proceeds only through the pathway with
the minimum energy barrier).
This change in the statistics may lead
to significant change in the branching ratios between
fission channels and to the alteration of the predominant
channel. 

For example, symmetric
$Na_{10}^{2+} \to 2Na_{5}^{+}$
and asymmetric $Na_{10}^{2+} \to Na_{7}^{+} + Na_3^{+}$
channels have very similar fission barriers, 
and one would expect that these two processes
would occur with similar probabilities.
The difference in the barriers heights of 0.01~eV
(see Table \ref{Na10bdTab})
would lead to the branching ratio between the
symmetric and asymmetric channels approximately equal to 2:3
at room temperature and 1:1 at higher temperatures.
If one adopts the statistics based on the binominal law,
the symmetric channel becomes the preferred fission channel,
since there are 252 combinations of five atoms out of ten and
only 120 combinations of three atoms out of ten.
Therefore, using the binominal statistics leads one to
conclude that the symmetric channel should prevail
in experimental mass spectra.

If, however, different combinations of atoms in the
parent cluster are considered non-equivalent, then
one needs to identify carefully the combinations of
atoms which provide the lowest separation barrier and
to count the numbers of such combinations.
In the asymmetric channel the lowest barrier is obtained
when an atom from a vertex and two neighbouring atoms
from the middles of the edges which cross at the vertex
are removed. There are totally 12 such combinations of atoms.
In the symmetric channel four atoms are removed
from the top of the pyramid and one from the middle
of the edge in the base of the pyramid. There are 12
such combinations, too. Hence, the statistics
which accounts for the cluster geometry leads
one to conclude that both channels statistically are
equally probable, so the branching ratio is governed by
the heights of the barriers only, i.e. the asymmetric channel
prevails.

The statistics based on the binominal law is more adequate
for high temperatures when a cluster is melted and does not
possess a certain structure\footnote{Melting temperature
of bulk sodium is 371~K, while for clusters it was estimated
to be about 300-400~K\cite{Manninen_PRL98,Andreoni_JCP91}}. 
To the contrary, 
at low temperatures, when the cluster structure is important,
the alternative statistics is more appropriate.

One can make an important conclusion based on these considerations
regarding the particular case of fission
of the $Na_{10}^{2+}$ cluster. Namely, at high temperatures
the symmetric channel is the preferred fission channel,
while at low temperatures when the cluster structure comes
into play the asymmetric channel slightly dominates.

\subsection{Fission dynamics: fission pathways and rearrangement during fission}

The potential energy surface approach
allows one to study the dynamics of the fission process. 
Having calculated the multi-dimensional potential energy surface
for $Na_{10}^{2+}$ and $Na_{18}^{2+}$ clusters
we were able to determine the optimal pathways
for all  the considered fission channels, i.e.
we have determined the coordinates of all atoms
which minimize the total cluster energy 
on each step of the simulation 
for each fission channel.

We have observed that
often fission barrier can be greatly lowered
by drawing the fission pathway via a local minimum
on the potential energy surface.
In other words, fission can proceed via formation 
of intermediate isomers.
This requires extra rearrangement 
of the cluster structure
as compared to a more straightforward fission pathway.

One can distinguish between two main kinds of
such rearrangement.\cite{Rearrangement_JPB04} 
The first one is rearrangement
of the cluster structure without significant change
in distance between the centers of mass of the prospective
fragments. This type of rearrangement takes place
before actual separation of the fragments begins
and it is, in fact, a transition to another isomer
state of the parent cluster.
The second type of rearrangement is characterized
by existence of a super-molecule-like intermediate state
in which the fragments are sufficiently separated
from each other, but connected by a 
"neck".\cite{Rearrangement_JPB04,Montag_PRB95,LandmanBrechignac_PRL94,Landman_PRL91}
A similar necking phenomenon 
is known for nuclear fission \cite{EisembergGreiner_85}.
Necking 
is responsible for a double humped form of
the fission barrier.
In dynamical simulations 
necking can be observed
as an elongation of the cluster shape
during fission.\cite{Guet_PRL01,BlaistenBarojas_PRB98}

\begin{figure}[ht]
\begin{center}
\includegraphics[scale=1]{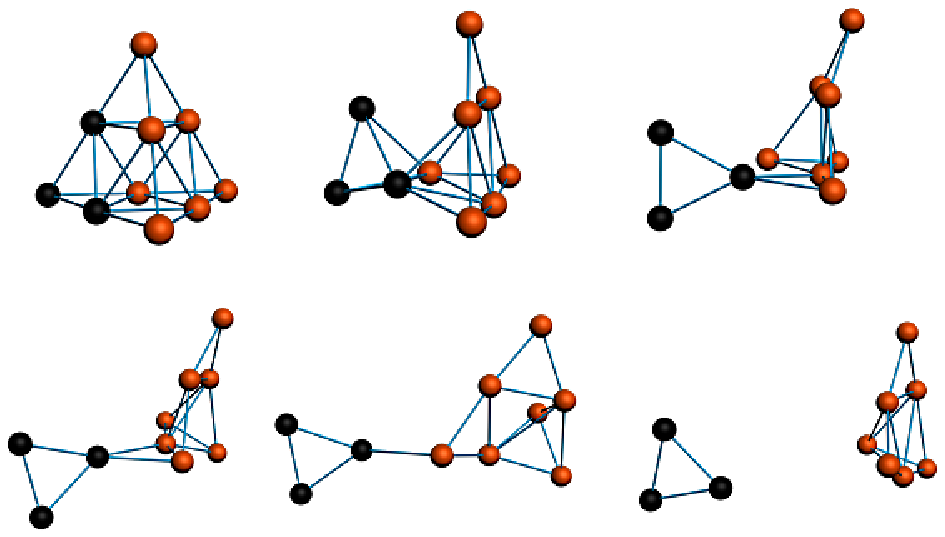}
\end{center}
\caption{(Color online) Rearrangement of the second kind of the cluster structure
during the fission process $Na_{10}^{2+} \to Na_7^+ + Na_3^+$
(fissioning atoms are shown in black).
First row, from left to right:
the ground state of the parent cluster (isomer $T_d^{(1)}$);
two atoms are pulled out of the middles of the edges;
the first transitional state corresponding to 
the top of the first hump on the fission barrier.
Second row, from left to right:
metastable super-molecule-like intermediate state corresponding
to the outer well of the fission barrier,
note the "necking" between the two fragments;
the second transitional state corresponding to
the top of the second hump on the fission barrier,
note how stretched the bonds are;
two fragments are drifting away from each other.}
\label{rearrangement}
\end{figure}

For the $Na_{10}^{2+}$ cluster, the rearrangement of the second
type takes place in the three out of five fission channels
which can be easily recognized by the double humped form
of the corresponding fission barriers in Figure \ref{Na10barriers}.
A typical example of such rearrangement is presented
in Figure \ref{rearrangement}. 
The key point in this process is the transition from
the first transitional state (the most right geometry in the
first row) to the intermediate state (the most left figure
in the second row). Exactly this transition results in
significant lowering of the fission barrier,
down to 0.34 eV from 0.48 eV in the situation when the 
atoms are continued to be pulled out from the cluster
and the intermediate state is not allowed to form.
The intermediate state is a super-molecule-like extended
structure with a neck-like connection between its parts.
The further fissioning of this intermediate structure 
is a simple stretching the neck and eventual complete
separation of the fragments.

Let us now return to the first type of rearrangement. 
This type of rearrangement implies that
there is a competition between one-step
(``direct'' fission) and two-step (fission 
via an intermediate
isomer state) processes. Therefore,
in order to make predictions concerning
the fission activation energy and fission pathway
for a given cluster isomer
one has first to find energy barrier for the direct fission
of the given isomer and then compare the height
of this barrier with the heights of the barriers
for transitions between given and several other low lying isomers,
and also with the heights of the fission barriers of those
isomers.

\begin{figure}[ht]
\begin{center}
\includegraphics[scale=0.5]{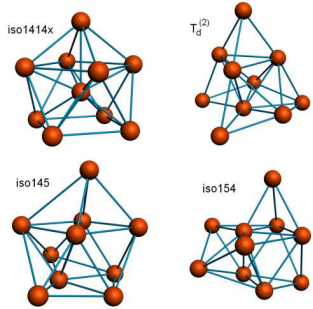}
\end{center}
\caption{(Color online) Four energetically favourable isomers of the $Na_{10}^{2+}$ cluster. 
A nickname for each of the isomers is given.
The geometry of the ground state isomer $T_d^{(1)}$
is shown in Figure \protect{\ref{Td}}.
Energies and symmetry groups of the isomers are summarized
in Table \protect{\ref{isoenergies}}.
}
\label{isogeometries}
\end{figure}

We present such analysis for the $Na_{10}^{2+}$ cluster. 
The energies of the several energetically favourable
isomers are summarized in Table \ref{isoenergies},
their geometries are shown in 
Figure \ref{isogeometries}. \footnote{Note, that the isomer with
$C_{4v}$ group of symmetry presented here is different
from the isomer shown in Figure \ref{isostructures}
and discussed in the previous section.}
The energy barrier heights for transitions between these
isomers are presented in Table \ref{isotransitions}.

\begin{table}
\caption{\label{isoenergies} 
Energies $E$ of the first several isomer states of $Na_{10}^{2+}$
cluster. Geometries of the isomers are shown in Figure \protect{\ref{isogeometries}}. 
The calculations are done with the use of
B3LYP exchange-correlation functional, molecular orbitals
are expanded in 6-31G(d) basis. The energies are measured
in atomic units, also differences with regard to the energy of
the ground state (in eV) are given.
}
\begin{ruledtabular}
\begin{tabular}{@{}lccc}
Isomer's nickname $\quad$ & Isomer's symmetry group$\quad$ & Isomer's energy $E$ $\quad$ & $E-E(T_d)$  \\
\hline
$T_d^{(1)}$ & $C_1$, distorted $T_d$ & -1622.6368 & 0.000 \\
iso1414x & $C_{4v}$ & -1622.6352 & 0.043 \\
$T_d^{(2)}$ & $C_1$, strongly distorted $T_d$ & -1622.6338 & 0.082 \\
iso145 & $C_{1}$ & -1622.6338 & 0.082 \\
iso154 & $C_{1}$ & -1622.6325 & 0.118 \\
\end{tabular} 
\end{ruledtabular}
\end{table}

\begin{table}
\caption{\label{isotransitions} 
Energy barrier heights for transitions between the first 
several low-lying isomer states of $Na_{10}^{2+}$
cluster. The calculations are done with the use of
B3LYP exchange-correlation functional, molecular orbitals
are expanded in 6-31G(d) basis. The barrier heights are measured
in eV.
}
\begin{ruledtabular}
\begin{tabular}{@{}lrrrrr}
Initial isomer / Final isomer &$\quad$ $T_d^{(1)}$ &$\quad$ iso1414x &$\quad$ $T_d^{(2)}$ &$\quad$iso145& $\quad$iso154 \\
\hline
$T_d^{(1)}$                    & --  &0.12 &0.092& 0.11& 0.13 \\
iso1414x                       &0.08 & --  &0.08 &0.08 &0.08 \\
$T_d^{(2)}$                    &0.01 &0.04 & --  &0.017&0.04 \\
iso145                         &0.03 &0.04 &0.017& --  &0.05 \\
iso154                         &0.01 &0.01 &0.01 &0.01 & -- \\
\end{tabular} 
\end{ruledtabular}
\end{table}

The analysis of the potential energy barriers for transitions
between different isomer states of the $Na_{10}^{2+}$ cluster
and of the potential energy barriers of fragmentation of the cluster
leads to an interesting fact.
It turns out that there are two fission pathways 
energetically equally probable
in the symmetric fission of the $Na_{10}^{2+}$ cluster,
$Na_{10}^{2+} \to 2 Na_{5}^+$.
Both of the pathways provide the minimum energy barrier (0.32 eV)
for the symmetric fragmentation of the parent cluster.
Both of the pathways involve rearrangement of the cluster's structure.
The energy barrier for the first pathway is shown in Figure
\ref{Na10barriers}. According to the first pathway,
the fission starts from separating the two
prospective fragments which later form a metastable super-molecule-like
structure which leads to a well in the middle of the fission barrier.
The second fission pathway constitutes an even more complicated,
two-stage process.
On the first stage the base of the pyramid in the parent cluster $T_d^{(1)}$
reshapes from the triangle into a non-planar pentagon with an atom
in its center forming a new isomer denoted $T_d^{(2)}$,
Figure \ref{isogeometries}.
After that the top four atoms in the pyramid and the atom from the center
of the pentagon can be removed from the rest of the cluster.
Again, this process proceeds via forming a necked-shaped
super-molecule-like structure.
For comparison, when the rearrangements are not allowed
and the chosen group of atoms is simply pulled out of the
parent cluster
the lowest possible energy barrier 
(for removal the top four atoms and an atom 
from a vertex of the base triangle) 
equals to 0.63 eV.

\begin{figure}[ht]
\begin{center}
\includegraphics[scale=1]{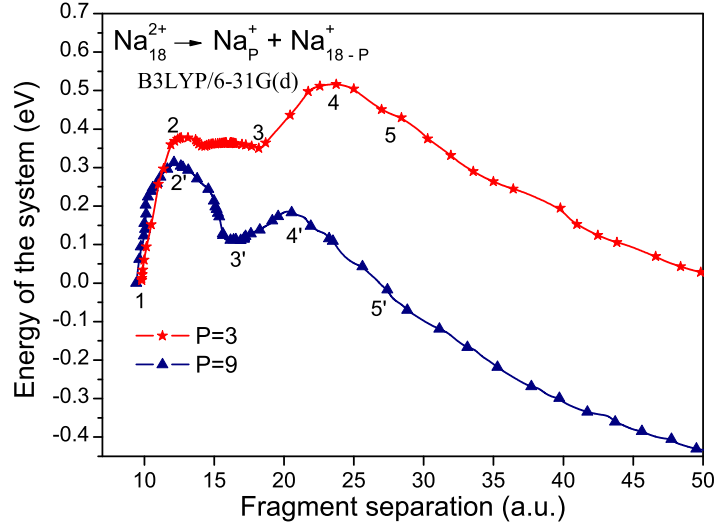}
\end{center}
\caption{(Color online) Fission barriers for the symmetric (triangles), $Na_{18}^{2+} \to 2 Na_9^+$,
and dominant asymmetric (stars), $Na_{18}^{2+} \to  Na_{15}^+ + Na_3^+$,
channels for $Na_{18}^{2+}$ cluster,
as functions
of distance between the centers of mass of the fragments.
Energy is measured from the energy of the ground $C_{5v}$ state of
the $Na_{18}^{2+}$ cluster. 
The geometries corresponding to different stages of the fission process 
in each channel are marked with subsequent numbers
and depicted in Figure \protect{\ref{Na18rearrangement}}.
}
\label{Na18barriers}
\end{figure}

\begin{figure}[ht]
\begin{center}
\includegraphics[scale=1]{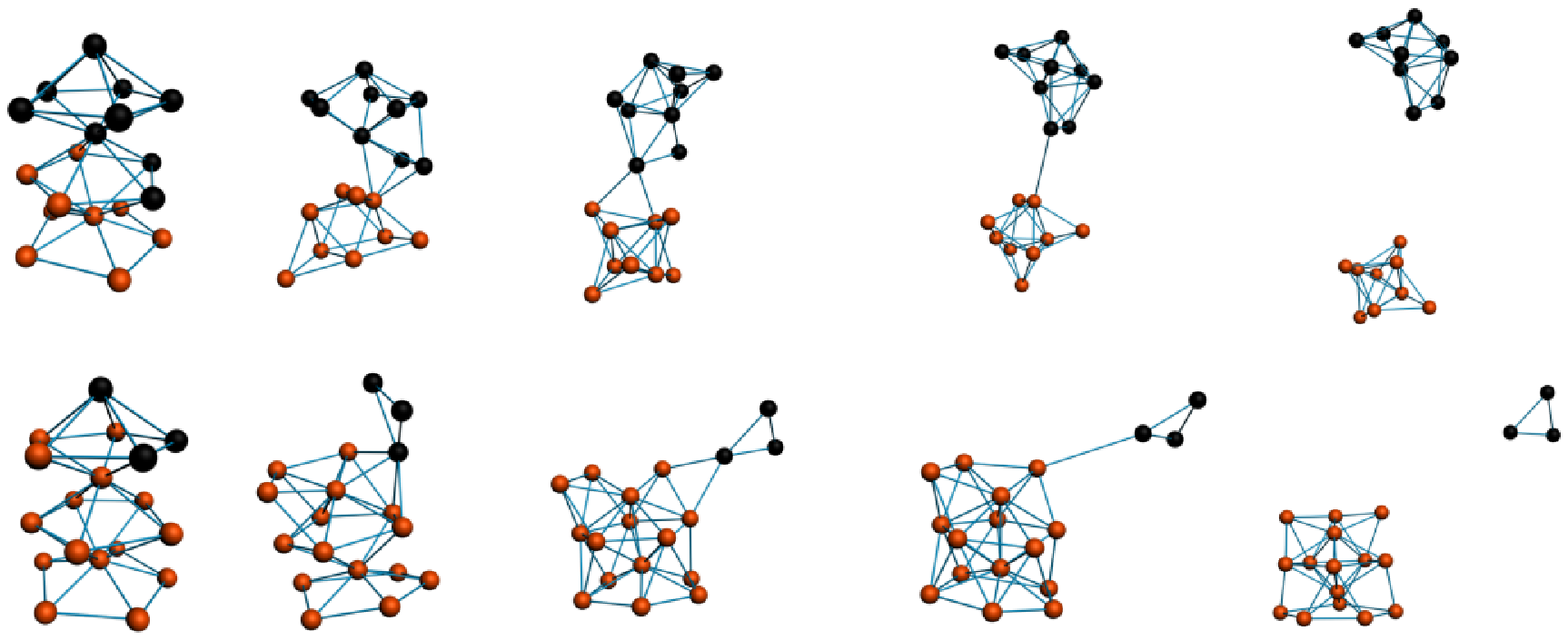}
\end{center}
\caption{(Color online) Rearrangement of the second kind of the cluster structure
during the fission processes $Na_{18}^{2+} \to 2 Na_9^+$ (upper row)
and $Na_{18}^{2+} \to  Na_{15}^+ + Na_3^+$ (lower row)
(fissioning atoms are shown in black).
Each subsequent geometry corresponds to a stage
of the fission process marked with a corresponding number
in Figure \protect{\ref{Na18barriers}}.}
\label{Na18rearrangement}
\end{figure}

The rearrangements of the cluster structure of the both types 
is a general feature of the metal cluster fission process.
They occur for other clusters and fission channels too.
As an illustration we plot in Figure \ref{Na18barriers} the fission barriers for fission
of $Na_{18}^{2+}$ cluster in symmetric and dominant asymmetric channels.
The geometries corresponding to different stages of the fission process 
in each channel are presented in Figure \ref{Na18rearrangement}.

To conclude, we have examined in detail fission of doubly charged
sodium clusters $Na_{10}^{2+}$.
Many new reference data are presented.
Three main conclusions can be drawn from our studies.
Firstly, geometry of the smaller fragment
and geometry of its immediate neighborhood in the parent cluster
(together with the electronic shell effects) 
play a leading role in defining the fission barrier height.
Secondly, 
rearrangement of the cluster structure in the course of fission
can lower the fission barriers significantly.
We distinguish two general types of rearrangement:
"necking" and fissioning via another low-lying isomer state
of the parent cluster.
And finally, accounting for geometrical structure of the cluster
leads to non-equivalence of different combinations of atoms
in the cluster which in turn 
affects calculating the cluster's entropy.

\acknowledgements
This work is supported by INTAS grant 03-51-6170
and by the
Russian Foundation for Basic Research
under the grant 03-02-18294-a.
A.G.L. is grateful to the Alexander von Humboldt Foundation
for financial support.
We acknowledge an access to computer cluster at
the Center for Scientific Computing of the
Johann Wolfgang Goethe-University
where the computations have been performed.

\end{document}